\documentclass[12pt,preprint]{aastex}
\begin{document}

\title{DIRECT Distances to Nearby Galaxies Using Detached Eclipsing
Binaries and Cepheids.  IX. Variables in the Field M31Y Discovered
with Image Subtraction}

\author{A. Z. Bonanos, K. Z. Stanek, D. D. Sasselov,
B. J. Mochejska\altaffilmark{1}}
\affil{Harvard-Smithsonian Center for Astrophysics, 60 Garden St.,
Cambridge, MA~02138}
\affil{\tt e-mail: abonanos@cfa.harvard.edu, kstanek@cfa.harvard.edu,
sasselov@cfa.harvard.edu, bmochejs@cfa.harvard.edu}

\author{L. M. Macri\altaffilmark{1}} 
\affil{National Optical Astronomy Observatories, 950 North Cherry
Avenue, Tucson, AZ 85719} 
\affil{\tt e-mail: lmacri@noao.edu}

\author{J. Kaluzny} \affil{Copernicus Astronomical Center, Bartycka
18, 00-716 Warszawa, Poland} 
\affil{\tt e-mail: jka@camk.edu.pl}
\altaffiltext{1}{Hubble Fellow}

\begin{abstract}

The DIRECT Project aims to obtain direct distances to two Local Group
galaxies, M31 and M33, which occupy a crucial position near the base
of the cosmological distance ladder. The first step is to search for
detached eclipsing binaries (DEBs) and Cepheids using 1 m class
telescopes to select good candidates, which will be followed up
spectroscopically on 6.5-10 m class telescopes. In this ninth paper we
present a catalog of variable stars discovered with image subtraction
in field M31Y [$\alpha=10.\!\!\arcdeg97, \delta=41.\!\!\arcdeg69, {\rm
J2000.0}$]. The data were obtained with the FLWO 1.2 m telescope on 25
nights, over a period of 6 months. In our search covering
$22\arcmin\times22\arcmin$ we discovered 41 eclipsing binaries, 126
Cepheids, and 97 other periodic or nonperiodic variables, including a
LBV candidate, a nova and a galactic CV. The catalog of variables, as
well as their photometry and finding charts, is available
electronically via {\tt anonymous ftp} and the {\tt World Wide
Web}. The complete set of the CCD frames is available upon request.

\end{abstract}

\section{Introduction}

Starting in 1996 we undertook a long term project, DIRECT
(i.e. ``direct distances''), to obtain the distances to two important
galaxies in the cosmological distance ladder, M31 and M33. These
``direct'' distances will be obtained by determining the distance to
Cepheids using the Baade-Wesselink method and by measuring the
absolute distance to detached eclipsing binaries (DEBs).

M31 and M33 are the stepping stones to most of our current effort to
understand the evolving universe at large scales. \citet{Wal03}
stresses the importance of M31: as a spiral galaxy, it is a more
suitable local galaxy for calibrating the distance scale than the LMC,
even though it is not as easy to observe. The difficulties include the
large angular extent of M31 on the sky, the variable (internal)
reddening and crowding. M31 and M33 also constrain population
synthesis models for early galaxy formation and evolution and provide
ample data for the stellar luminosity calibration. There is one simple
requirement for all this---accurate distances. These distances are now
known to no better than 10-15\%, as there are discrepancies of
$0.2-0.3\;{\rm mag}$ between various distance indicators.

DEBs have the potential to establish distances to M31 and M33 with an
unprecedented accuracy of 5\%. Detached eclipsing binaries \citep[for
reviews see][]{And91,Pac97} offer a single step distance determination
to nearby galaxies and may therefore provide an accurate zero point
calibration of various distance indicators -- a major step towards
very accurate determination of the Hubble constant, presently an
important but daunting problem for astrophysicists. In the last few
years, DEBs have been used to obtain accurate distance estimates to
the Large Magellanic Cloud \citep[e.g.][]{Gui98,Fit03} and the Small
Magellanic Cloud \citep{Har03}. Distances to individual DEBs in these
papers are claimed to be accurate to 5-10\%.

Detached eclipsing binaries have yet to be used as distance indicators
to M31 and M33. The DIRECT project has begun a massive search for
periodic variables, discovering so far 4 good DEBs, now that
large-format CCD detectors are available and that CPUs are
inexpensive. These DEBs will be spectroscopically followed-up with the
powerful 6.5-10 meter telescopes.

The study of Cepheids in M31 began with Hubble \citep{Hub29}. Later
Baade's photographic plates of fields I-IV \citep{Gap62, Baa63, Baa65}
and Magnier's CCD survey \citep{Mag97} led to the discovery of a few
hundred Cepheids. \citet{Wel86} obtained infrared photometry of
several Cepheids in Baade's fields. \citet{Fre90} obtained CCD
observations of known Cepheids in M31 to determine more accurate
periods and study metallicity effects. Recently, \citet{Jos03} have
obtained R and I band observations of a $13'\times 13'$ region in the
disk of M31 galaxy and derived a Cepheid P-L distance. However, the
existing photometry of M31 is sparse and does not provide a good basis
for obtaining direct Baade-Wesselink distances \citep[see,
e.g.][]{Kro97} to Cepheids---the need for new digital photometry has
been long overdue.

As the first step of the DIRECT project we have searched for DEBs and
new Cepheids in the M31 and M33 galaxies. We have analyzed five
$11\arcmin\times11\arcmin$ fields in M31, A-D and F \citep[][hereafter
Papers I, II, III, IV, V]{Kal98,Sta98a,Sta99,Kal99,Moc99}. A total of
410 variables, mostly new, were found: 48 eclipsing binaries, 206
Cepheids and 156 other periodic, possible long-period or non-periodic
variables. We have analyzed two fields in M33, A and B (Macri et
al. 2001a; hereafter Paper VI) and found 544 variables: 47 eclipsing
binaries, 251 Cepheids and 246 other variables. Follow up observations
of fields M33A and M33B produced 280 and 612 new variables,
respectively \citep[][hereafter Papers VII,
VIII]{Moc01a,Moc01b}. Variables from two more DIRECT fields, one in
M31 and the other in M33, remain to be reported.

In this paper, ninth in the series, we present a catalog of variable
stars found in field M31Y. The paper is organized as follows: Section
2 provides a description of the observations. The data reduction
procedure, calibration and astrometry is outlined in Section 3. The
catalog of variable stars is presented in Section 4, followed by a
brief discussion in Section 5.

\section{Observations}

\begin{figure}   
\plotone{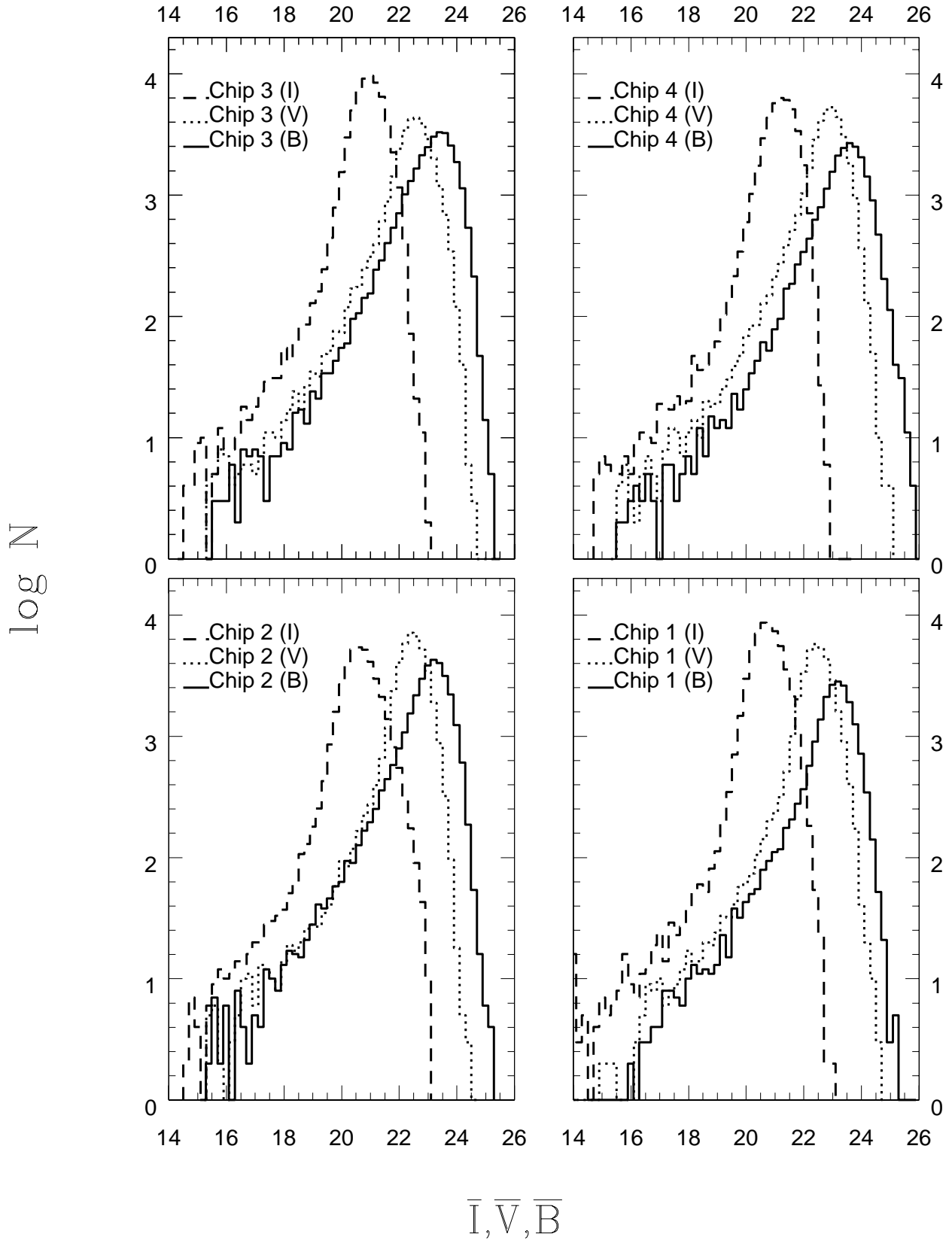}
\caption{Distribution of stars in each filter and chip, showing the depth
our survey is complete to.}
\label{completeness}
\end{figure}

The data for M31 field Y was taken with the 1.2m telescope at the Fred
Lawrence Whipple Observatory on Mount Hopkins, Arizona, between July
19th, 1999 and January 2nd, 2000, over 25 nights. We used the
``4Shooter'' camera \citep{Sze03}, with 4~thinned and AR-coated Loral
2048$^2$ pixel CCDs. The pixels are 15 microns in size and map to
$0.33\arcsec$ per pixel on the focal plane, making each image
$11\arcmin$ on the side. The camera was centered at
$\alpha=10.\!\!\arcdeg97, \delta=41.\!\!\arcdeg69, {\rm J2000.0}$,
which is approximately $0.\!\!\arcdeg48$ or 6.6 kpc from the nucleus,
assuming a distance to M31 of 784 Kpc \citep{Sta98b}. The data
consists of 126 $\times$ 900 $s$ exposures in the $V$ filter, 21
$\times$ 1200 $s$ exposures in the $B$ filter and 36 $\times$ 600 $s$
exposures in the $I$ filter. The median value of the seeing in $V$ was
$1.7\arcsec$. The field was observed through airmasses ranging from
1.02 to 2.11, with the median being 1.15. The completeness of our
photometry starts to drop rapidly at about 20.5 in $I$, 22.5 in $V$
and 23.5 in $B$, as shown in the distribution of stars in
Figure~\ref{completeness}. On two photometric nights of the run,
several images of standard \citet{Lan92} fields were taken.

\section{Data Reduction, Calibration and Astrometry}

Preliminary processing of the data was performed with standard
routines in the IRAF~\footnote{IRAF is distributed by the National
Optical Astronomy Observatories, which are operated by the Association
of Universities for Research in Astronomy, Inc., under cooperative
agreement with the NSF.} CCDPROC package. The photometry for the
variable stars was extracted using the ISIS image subtraction package
\citep{Ala98,Ala00} from the $V$-band data. 

The ISIS reduction procedure consists of several steps. Initially, all
the frames are transformed to a common coordinate grid. Next, a
reference image is created by stacking several frames with the best
seeing. For each frame, the reference image is convolved with a kernel
to match its PSF and then subtracted. On the subtracted images, the
constant stars will cancel out, and only the signal from variable
stars should remain. A median image is constructed of all the
subtracted images, and the variable stars are identified as bright
peaks on it. Finally, profile photometry is extracted from the
subtracted images. Paper VII describes this procedure in more detail.

\subsection{Photometric Calibration and Astrometry}

During each of the photometric nights of October 11 and November 3,
1999, we observed 7 \citet{Lan92} fields in the $BVI$ filters at air
masses ranging from 1.12 to 1.97 in $I$, 1.12 to 2.00 in $V$ and 1.13
to 2.05 in $B$. The transformation from the instrumental to the
standard system was derived for each chip in the following form:

\begin{eqnarray*}
b = B + \rm \chi_{b} + \rm \xi_{b} \cdot(B-V) + \rm \kappa_{b} \cdot X\\
v = V + \rm \chi_{v1} + \rm \xi_{v1} \cdot(B-V) + \rm \kappa_{v1} \cdot X\\
v = V + \rm \chi_{v2} + \rm \xi_{v2} \cdot(V-I) + \rm \kappa_{v2} \cdot X\\
i = I + \rm \chi_{i} + \rm \xi_{i} \cdot(V-I) + \rm \kappa_{i} \cdot X\\
\end{eqnarray*}

\noindent where lowercase letters correspond to the instrumental
magnitudes, uppercase letters to standard magnitudes and X is the
airmass. The values of the zeropoint ($\chi$), color ($\xi$) and
airmass coefficients ($\kappa$) are given in
Table~\ref{tab:coeff}. The zeropoints for the two nights agree to 0.02
mag. Since the color coefficients are small, we took $B-V=V-I=1$ when
transforming the magnitudes of our stars, which is approximately the
color of a Cepheid.

\begin{deluxetable}{lcccc}
\footnotesize \tablewidth{11.cm} \tablecaption{Transformation
coefficients for each chip.\label{tab:coeff}} \tablecolumns{4} 
\tablehead{ \colhead{Instrumental Mag.} &  \colhead{$\chi$} &
\colhead{$\xi$} & \colhead{$\kappa$} }
\startdata
b$_{1}$(B-V).....      & $-5.25$ & $-$0.025 & 0.216  \\
v$_{1}$(B-V).....      & $-5.30$ &  0.039 & 0.145  \\
v$_{1}$(V-I).....      & $-5.30$ &  0.035 & 0.146  \\
i$_{1}$(V-I).....      & $-4.47$ &  0.002 & 0.088  \\ 
\tableline		              	      
b$_{2}$(B-V).....      & $-5.00$ & $-$0.056 & 0.192  \\
v$_{2}$(B-V).....      & $-4.92$ &  0.037 & 0.136  \\
v$_{2}$(V-I).....      & $-4.92$ &  0.032 & 0.137  \\
i$_{2}$(V-I).....      & $-4.37$ &  0.012 & 0.079  \\
\tableline				              	      
b$_{3}$(B-V).....      & $-5.48$ & $-$0.033 & 0.220  \\
v$_{3}$(B-V).....      & $-5.47$ &  0.047 & 0.136  \\
v$_{3}$(V-I).....      & $-5.47$ &  0.041 & 0.136  \\
i$_{3}$(V-I).....      & $-4.69$ &  0.040 & 0.070  \\
\tableline				              	      
b$_{4}$(B-V).....      & $-5.16$ & $-$0.046 & 0.225  \\
v$_{4}$(B-V).....      & $-5.28$ &  0.038 & 0.146  \\
v$_{4}$(V-I).....      & $-5.29$ &  0.033 & 0.146  \\
i$_{4}$(V-I).....      & $-4.61$ &  0.009 & 0.066  \\
\enddata
\end{deluxetable}

\begin{figure}   
\plotone{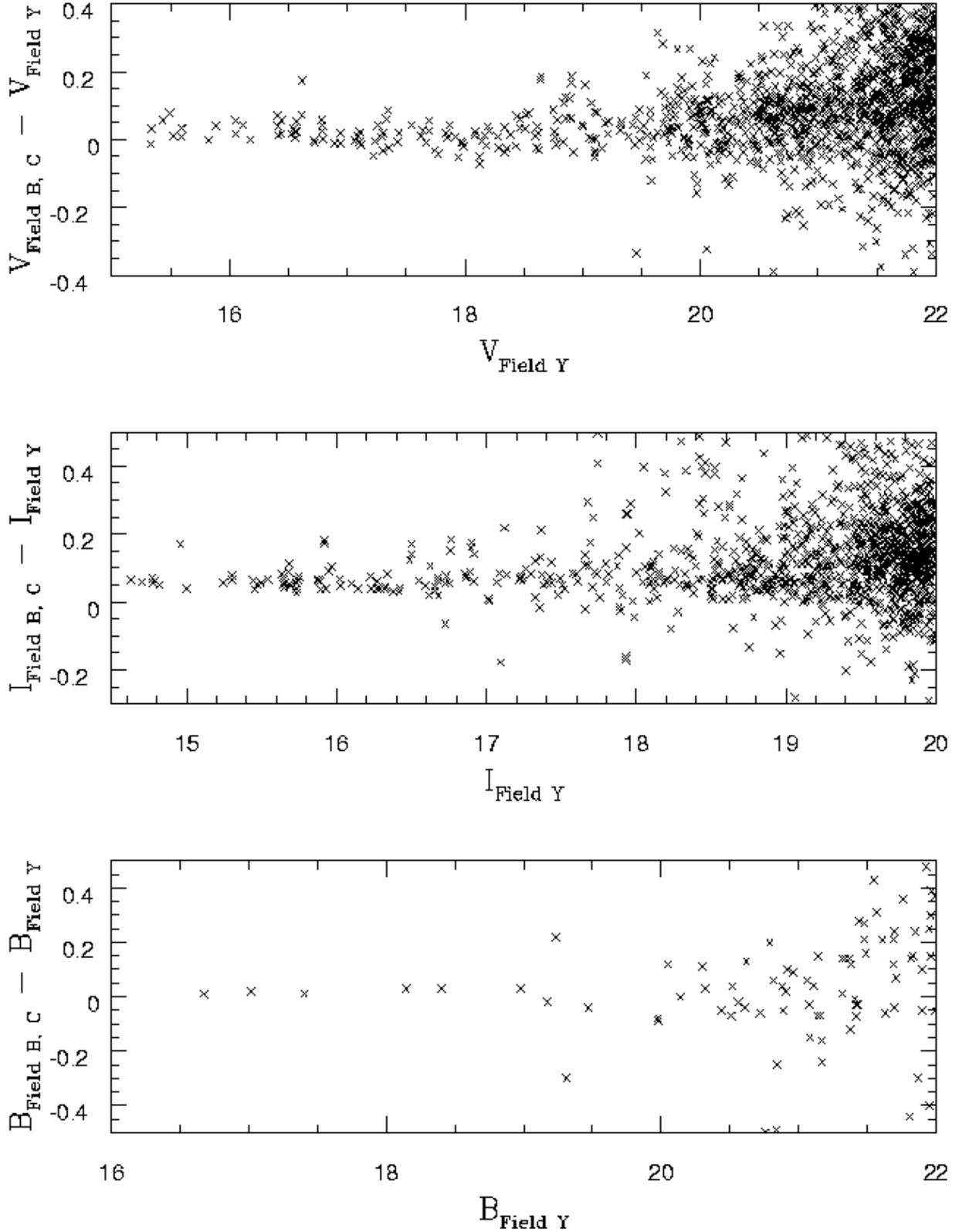}
\caption{Comparison of our photometry with overlapping fields B and C
(Paper I; Paper III) in $V$, $I$ and $B$ bands. The median offsets
found in each are 0.02, 0.06 and 0.03 mag respectively.}
\label{BCvsY}
\end{figure}

We compared our photometry with overlapping DIRECT Field M31B and M31C
photometry (Paper I and Paper III) and also with Magnier's photometry
\citep{Mag97}. The median magnitude offsets with DIRECT Field B and C
photometry are 0.02 mag in $V$, 0.06 mag in $I$ and 0.03 mag in $B$,
as shown in Figure~\ref{BCvsY}. We compared 2299 stars in $V$, 2383 in
$I$ and 320 in $B$. Our $I$-band photometry is slightly brighter (0.06
mag) than in Paper I, while the photometry of \citet{Jos03} is fainter
by 0.13 magnitudes. We believe that the $I$ band magnitudes
of \citet{Jos03} are off.

\begin{figure}   
\plotone{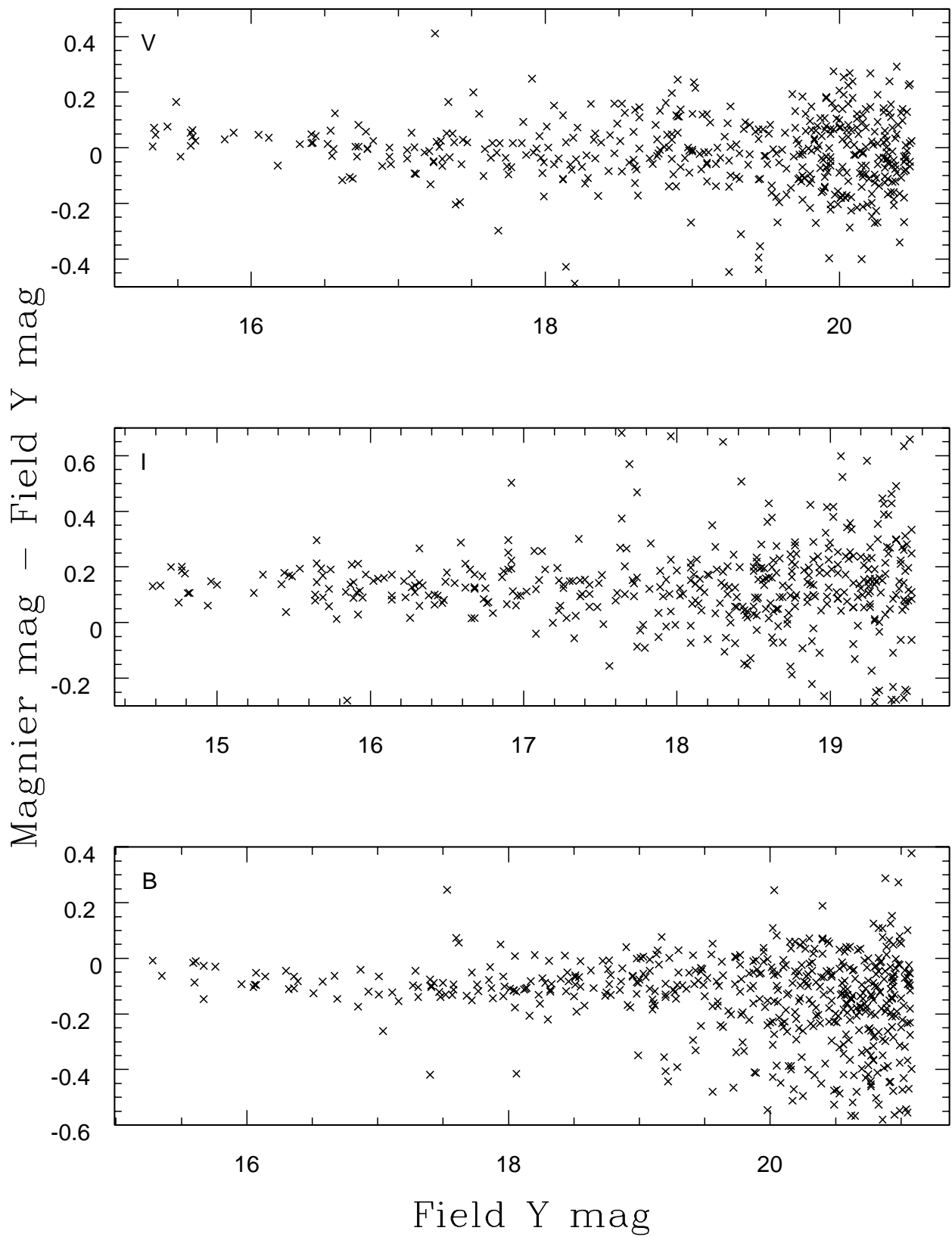}
\caption{Comparison of our Chip 2 photometry with the Magnier catalog
\citep{Mag97} in $V$, $I$ and $B$ bands. The median offsets found in
each are 0.01, 0.133 and $-$0.101 mag respectively.}
\label{Magniercomp}
\end{figure}

Figure~\ref{Magniercomp} shows the comparison of our photometry with
Magnier's photometry for Chip~2, which is representative of the other
chips as well. The average $V$ magnitude differences between Magnier's
photometry and ours, for stars brighter than 18th magnitude, for
Chips~1~-~4 were: $-$0.06, 0.01, $-$0.03 and $-$0.14 mag, by comparing
1560, 3783, 747 and 206 matching stars, respectively. In $I$ the
offset with Magnier is $\sim 0.1$ mag, while in $B$ the offset is $\sim
-0.2$ mag. Since there is good agreement with Paper I in these bands,
we believe that the Magnier $B$ and $I$ magnitudes are off and should
be used with caution.

Equatorial coordinates were determined for the $V$ star list. The
transformation from rectangular to equatorial coordinates was derived
using for Chips 1-4: 135, 117, 191 and 221 transformation stars
respectively with $V<20$ from the USNO-A2.0 \citep{Mon96} catalog. The
average difference between the catalog and the computed coordinates
for the transformation stars was less than $0.\arcsec3$ in RA and
$0.\arcsec3$ in Dec. We also compared the astrometry to Magnier's
catalog and found 262, 313, 307 and 72 matches for Chips 1-4, having a
median offset $<0.\arcsec4$. We use these derived equatorial
coordinates to name the variables, adopting the convention after
\citet{Mac01} based on the J2000.0 equatorial coordinates, in the
format: D31J$hhmmss.s$+$ddmmss.s$. The first three fields ($hhmmss.s$)
correspond to RA expressed in hours, the last three ($ddmmss.s$) to
Dec, expressed in degrees, separated by the declination sign.

\section{Catalog of Variables}

The preliminary classification process used, as described in Paper I,
classified the variable stars as eclipsing, Cepheids, or
miscellaneous. In order to obtain a clean Cepheid sample, we
reclassified Cepheids with highly discrepant colors on a CMD (shown in
Figure~\ref{cmd}) as other periodic variables. Also, extreme outliers
on the period-luminosity relation in $V$ and $I$ (see Figure~\ref{pl})
were reclassified as other periodic variables, as they most likely are
Type II Cepheids. Next, we present the catalog of light curves and
parameters of 264~variables.\footnote{The $BVI$ photometry and $V$
finding charts for all variables are available from the authors via
the {\tt anonymous ftp} on {\tt cfa-ftp.harvard.edu}, in {\tt
pub/kstanek/DIRECT} directory and can be also accessed through the
{\tt World Wide Web} at {\tt
http://cfa-www.harvard.edu/\~\/kstanek/DIRECT/}.}

\subsection{Eclipsing Binaries}
\begin{figure}   
\plotone{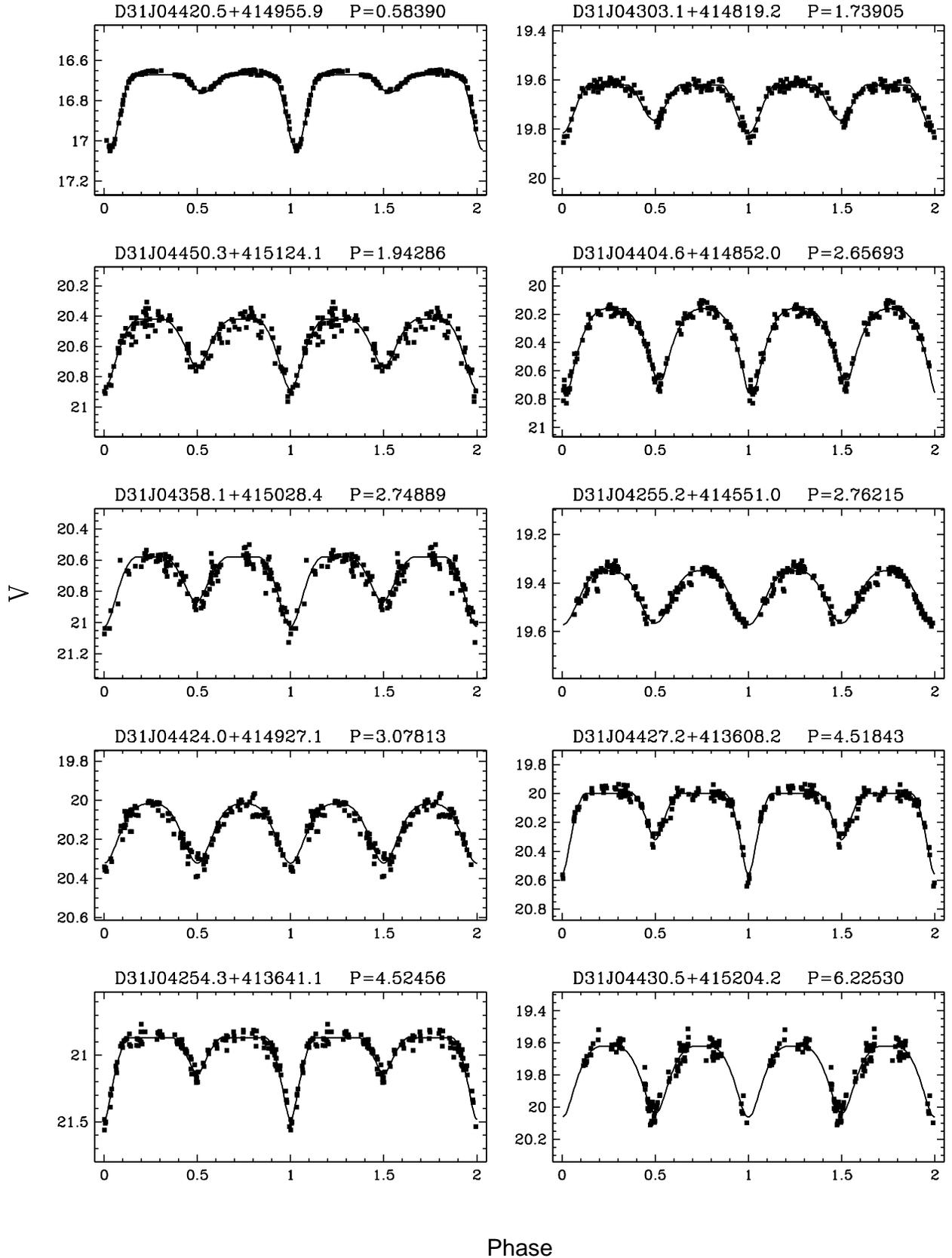}
\caption{Sample $V$-band light curves of EBs found in M31Y. The solid line is
the best fit model to the binary.}
\label{ebs}
\end{figure}

We found 41 EBs in field M31Y which are presented in. In
Table~\ref{tab:ebs} we list their name, period P, magnitudes
$B_{max}$, $V_{max}$ and $I_{max}$ of the system outside of the
eclipse, and the radii $R_{1}, R_{2}$ in the units of the orbital
separation. We also give the inclination angle $i$ of the binary orbit
to the line of sight and the eccentricity $e$. These values are
determined from a simple model of the eclipsing system, so they should
be treated only as reasonable estimates of the ``true'' values. The
table also includes the flux EBs, for which only the position and
period are given. Figure~\ref{ebs} presents the phased light curves of
10 sample EBs. Table~\ref{tab:ecl_lc} presents all the EB light
curves.

\subsection{Cepheids}
\begin{figure}   
\plotone{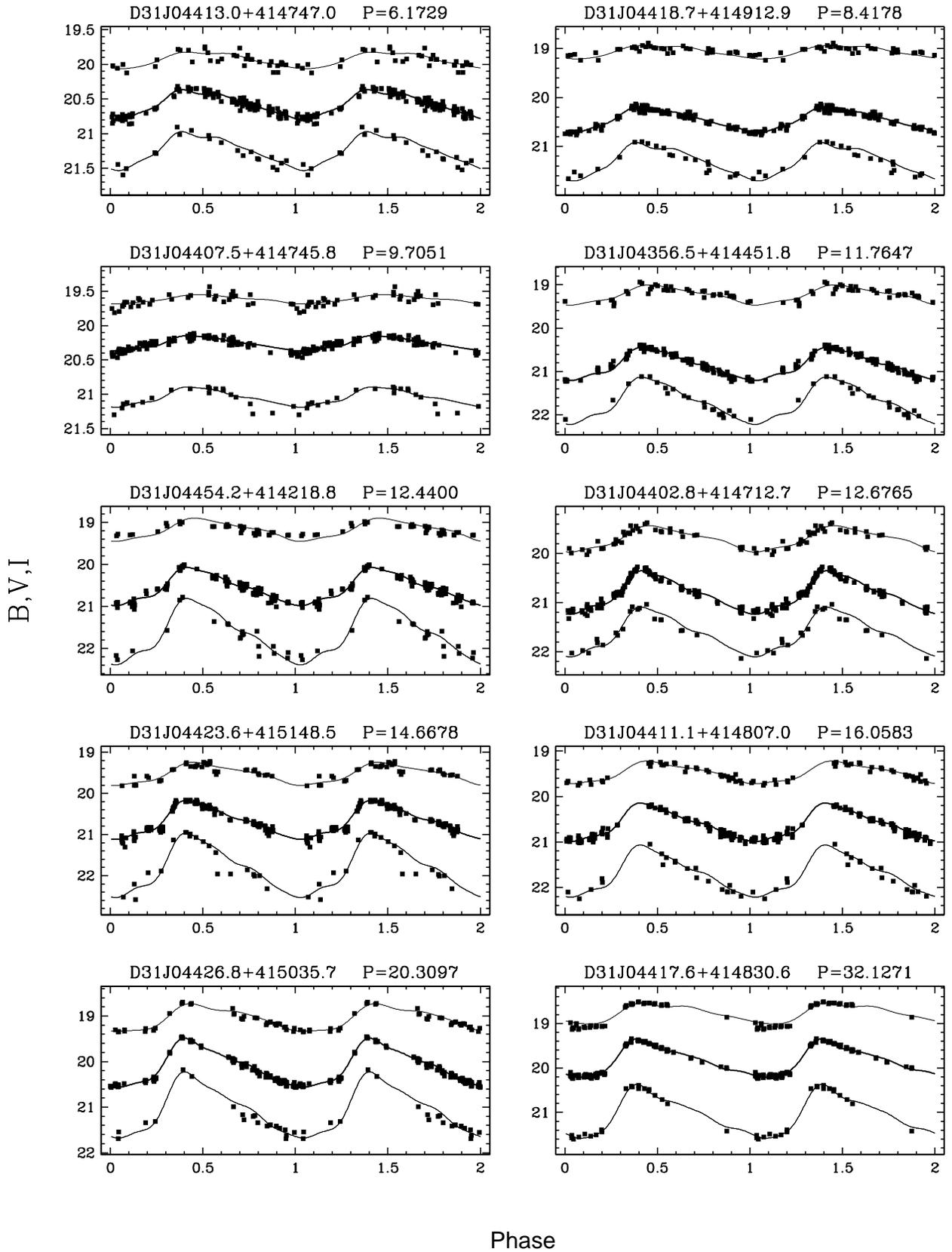}
\caption{Selected $B, V, I$ light curves for Cepheids in M31Y
Chip3. The solid line is the best-fit model.}
\label{ceph}
\end{figure}

A total of 126 Cepheids were found in field M31Y. In
Table~\ref{tab:ceph} we present their light curves and parameters,
ordered by increasing period. Specifically, the table lists the
Cepheid name, period $P$, $V$-band amplitude $A_{V}$ and flux-weighted
average magnitudes $\langle V\rangle$, $\langle I\rangle$ and $\langle
B\rangle$. For Cepheids with flux light curves, only the name and
period are given. There are 15 previously identified Cepheids, for
which we reference their discovery name. Figure~\ref{ceph} presents
the phased light curves of 10 sample Cepheids. Table~\ref{tab:ceph_lc}
presents all the Cepheid light curves.

\subsection{Other Periodic Variables}
\begin{figure}   
\plotone{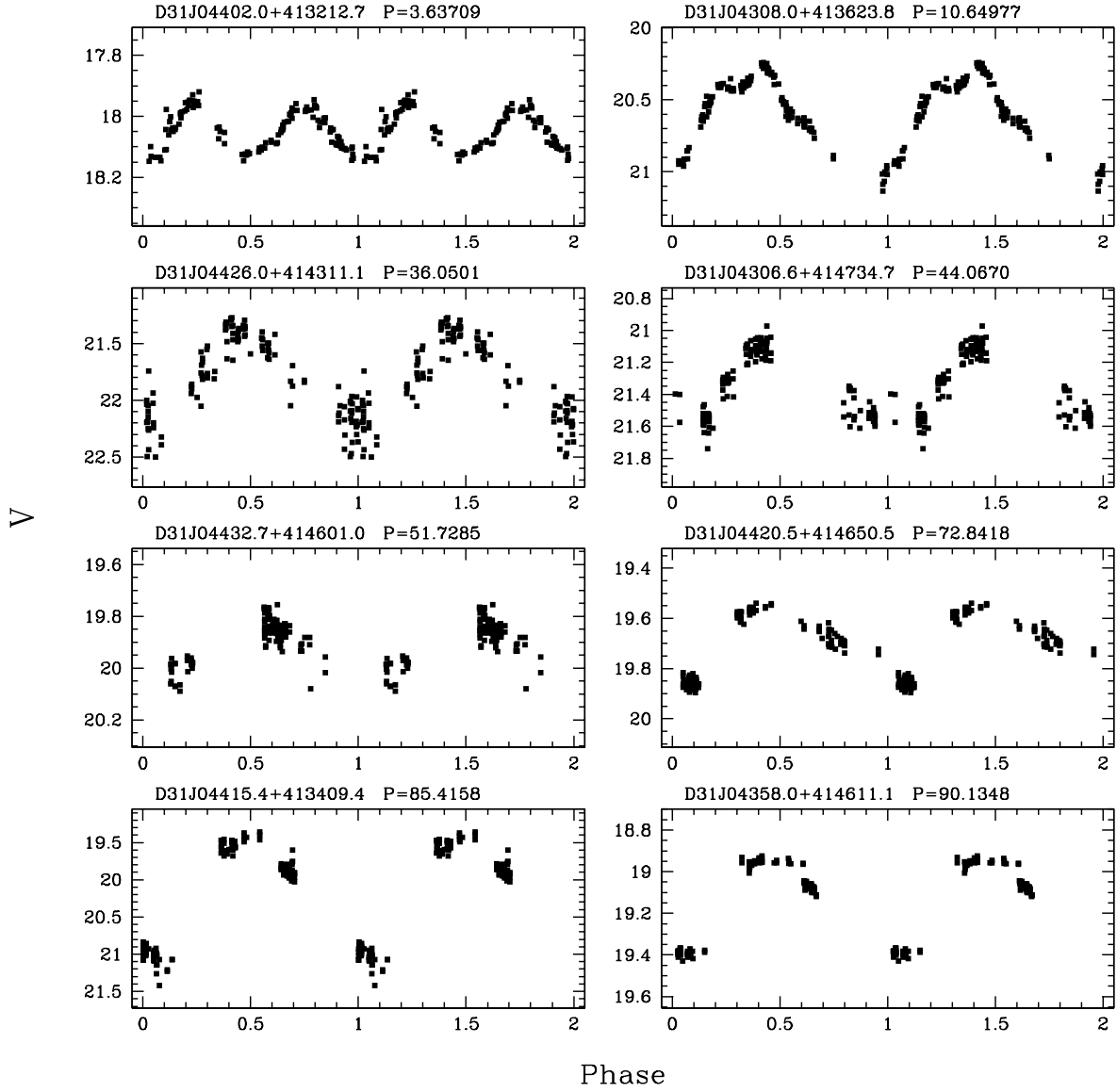}
\caption{Sample V-band light curves of eight other periodic variables.}
\label{oper}
\end{figure}

In Table~\ref{tab:per} we present parameters of 48 periodic variables,
ordered by increasing period. We list each variable's name, period $P$
and flux-weighted average magnitudes $\langle V\rangle$, $\langle
I\rangle$ and $\langle B\rangle$. Figure~\ref{oper} presents phased
light curves of eight sample periodic variables. Several of these are
Type II cepheids. Table~\ref{tab:per_lc} presents all the light curves of
other periodic variables.

\subsection{Miscellaneous Variables}
\begin{figure}   
\plotone{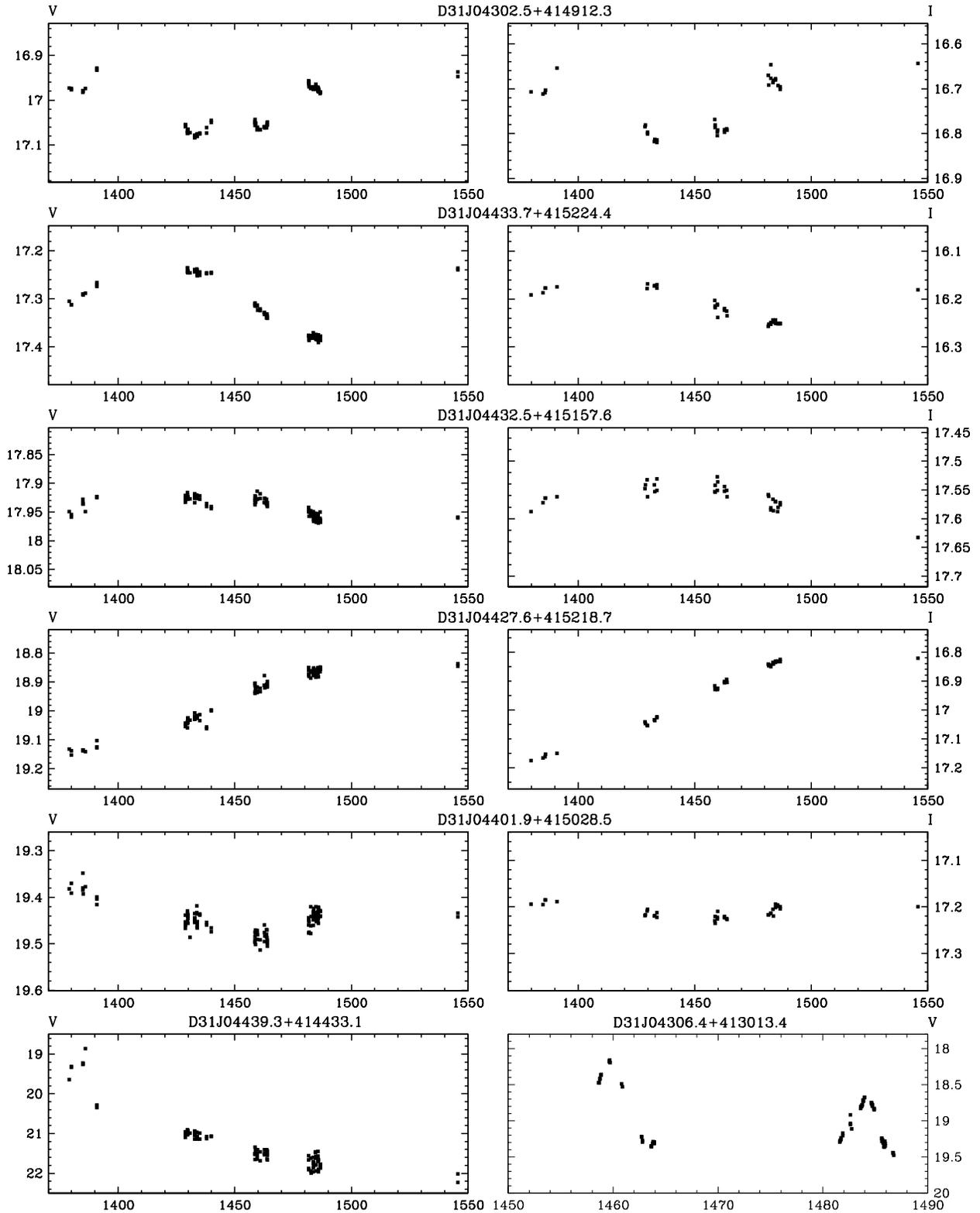}
\caption{Sample light curves of miscellaneous variables.}
\label{misc}
\end{figure}

In Table~\ref{tab:misc} we present parameters of 49 miscellaneous
variables, ordered by increasing RA. We list the name of each variable
and the average $V$, $I$ and $B$ magnitudes. Figure~\ref{misc}
presents $V$-band and $I$-band light curves for several miscellaneous
variables. Table~\ref{tab:misc_lc} presents all the light curves of
miscellaneous variables.

There are several interesting variables in this category worth
mentioning. D31J04439.3 +414433.1 is a possible nova, which dropped 3
magnitudes over 100 days in $V$. D31J04302.5 +414912.3 seems to be a
luminous blue variable (LBV). Finally, D31J04306.4+413013.4 is
probably a foreground cataclysmic variable, similar to Z Cam. Near
both maxima it has $B-V=0.24$ and $V-I=0.5$.

\section{Discussion}

\begin{figure}   
\plotone{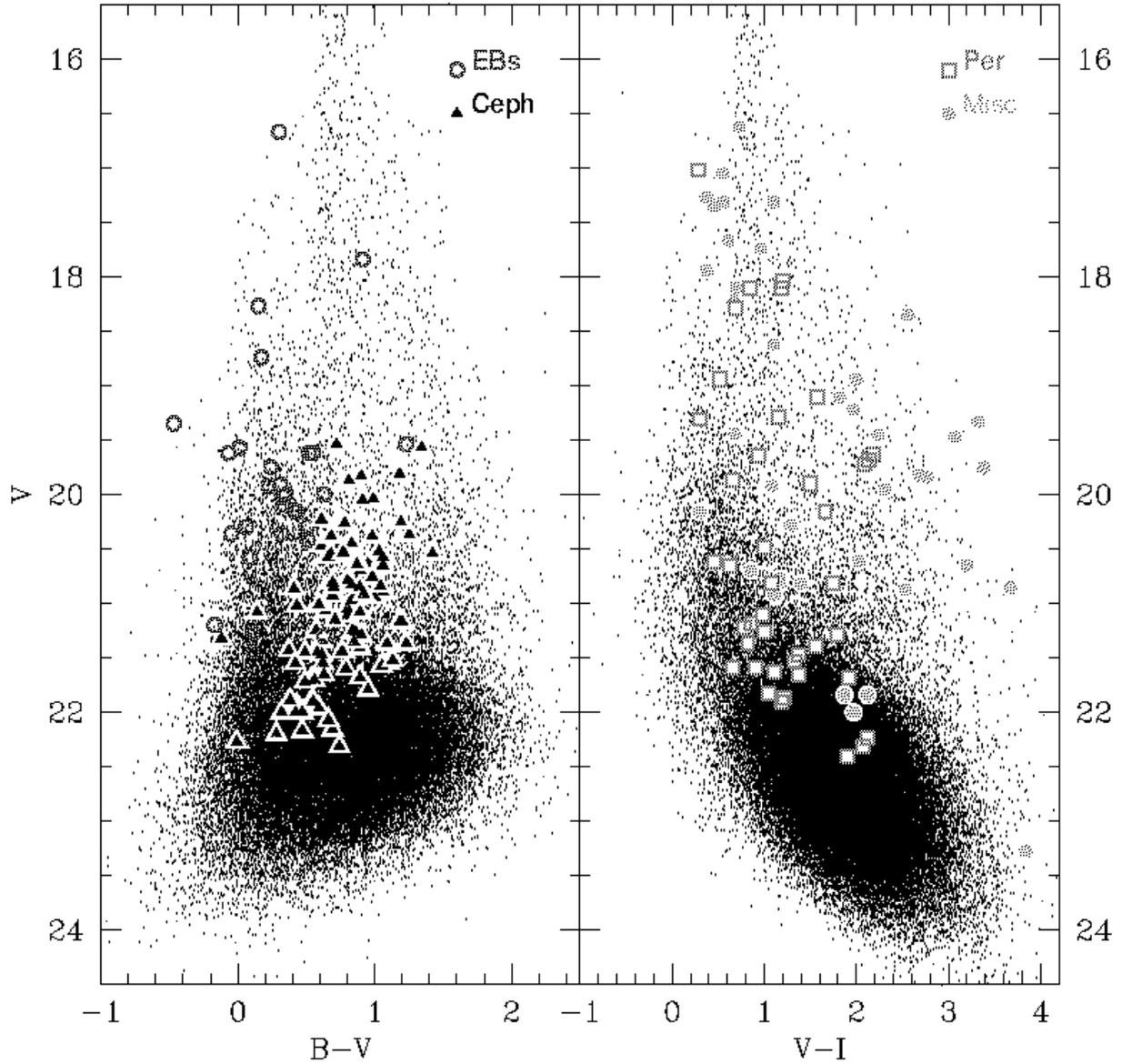}
\caption{The $V/B-V$ and $V/V-I$ CMD for stars in M31Y. The EBs are
marked with open circles, Cepheids with filled triangles on the right
panel; on the left panel the periodic variables are open squares and
the miscellaneous variables are marked with filled circles. [{\it See
the electronic edition of the Journal for a color version of this
figure.}]}
\label{cmd}
\end{figure}

In Figure~\ref{cmd} we plot the location of the variables on the
CMD. In the left panel, the positions of EBs (open circles) and
Cepheids (filled triangles) are shown on a $V/B-V$ CMD. In the right
panel the location of other periodic (open squares) and miscellaneous
variables (filled circles) are shown on a $V/V-I$ CMD. Most of the EBs
occupy the upper main sequence. Variable D31J04420.5+414955.9, with
$V=16.67$ and $B-V=0.3$, is most likely a foreground DEB, since it is
bright and has a period of only 0.58390 days, even though it is
projected onto the spiral arms. No extinction correction has been
applied, resulting in a spread of the EBs and Cepheids on the CMD.

\begin{figure}  
\plotone{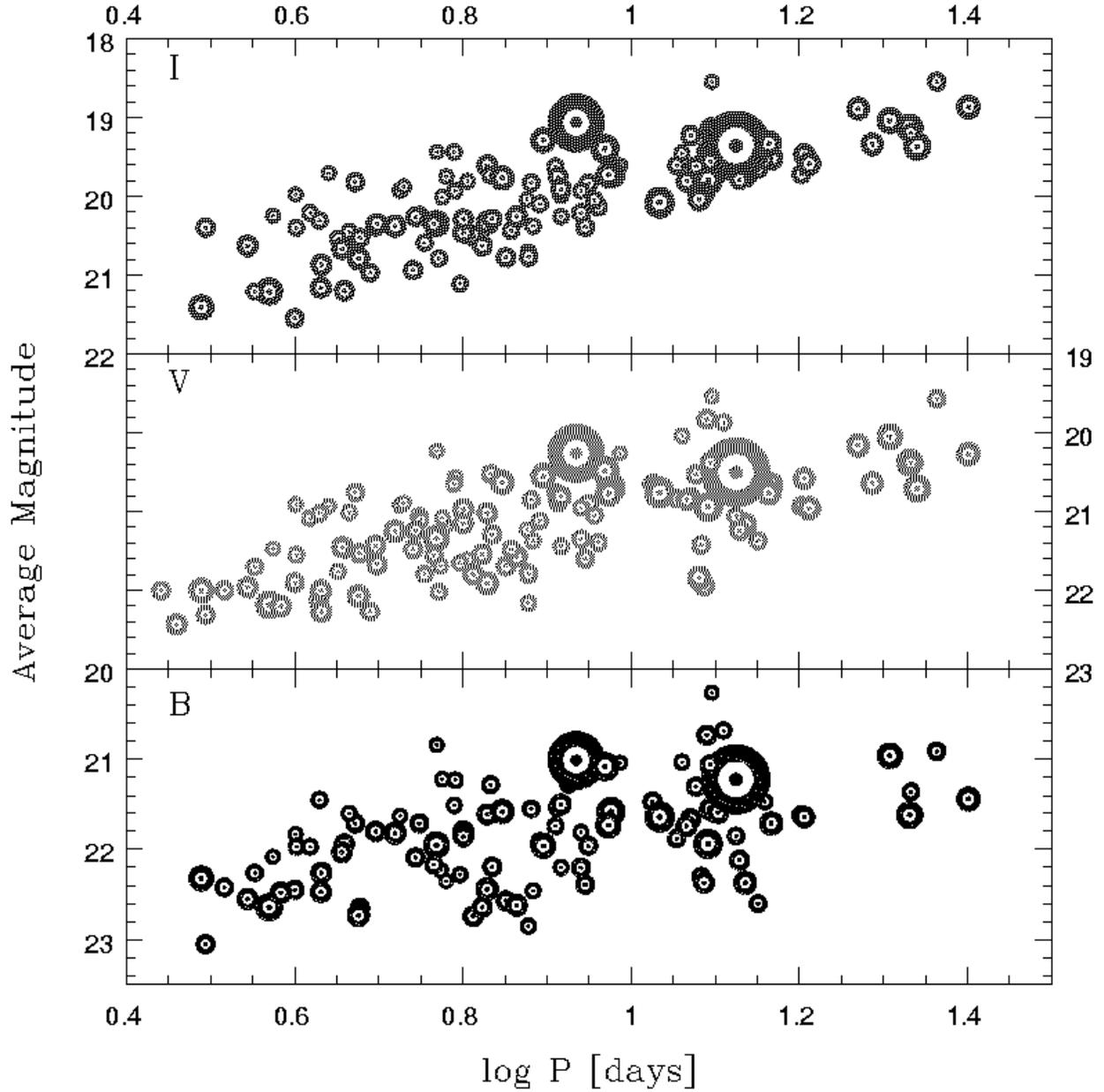}
\caption{PL relation for M31Y Cepheids in $I$, $V$ and $B$. The size
of the circle is proportional to the amplitude of the Cepheid in each
band. [{\it See the electronic edition of the Journal for a color
version of this figure.}]}
\label{pl}
\end{figure}

The Cepheid period-luminosity relation for the $I$, $V$ and $B$ bands
is shown in Figure~\ref{pl}. The size of the circles representing
Cepheids is proportional to their amplitude. The scatter in each plot
is due to extinction and errors in the period determination and the
photometry.

\begin{figure}  
\plotone{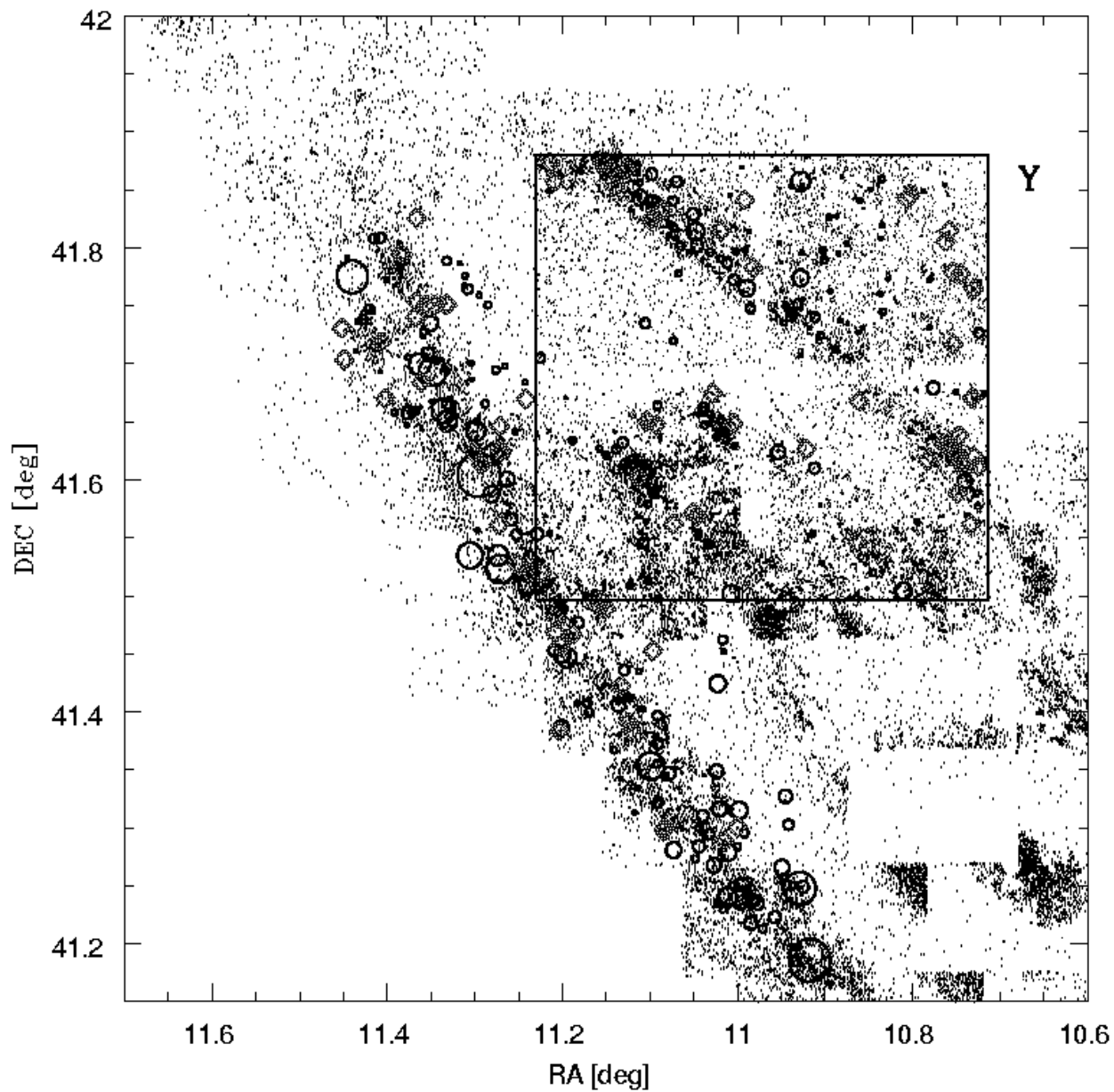}
\caption{The distribution of Cepheids and EBs in the northeastern part
of M31. Magnier's catalog \citep{Mag97} is used to outline the
galaxy. For Field Y, 14,000 of our stars ($V<21.8$) are also
overlayed. Circles represent the Cepheids and are proportional to the
period; diamonds represent the EBs. The lack of Cepheids in Field Y
with period $>25$ days, may suggest more recent star formation on side
of M31. [{\it See the electronic edition of the Journal for a color
version of this figure.}]}
\label{vardist}
\end{figure}

\begin{figure}  
\plotone{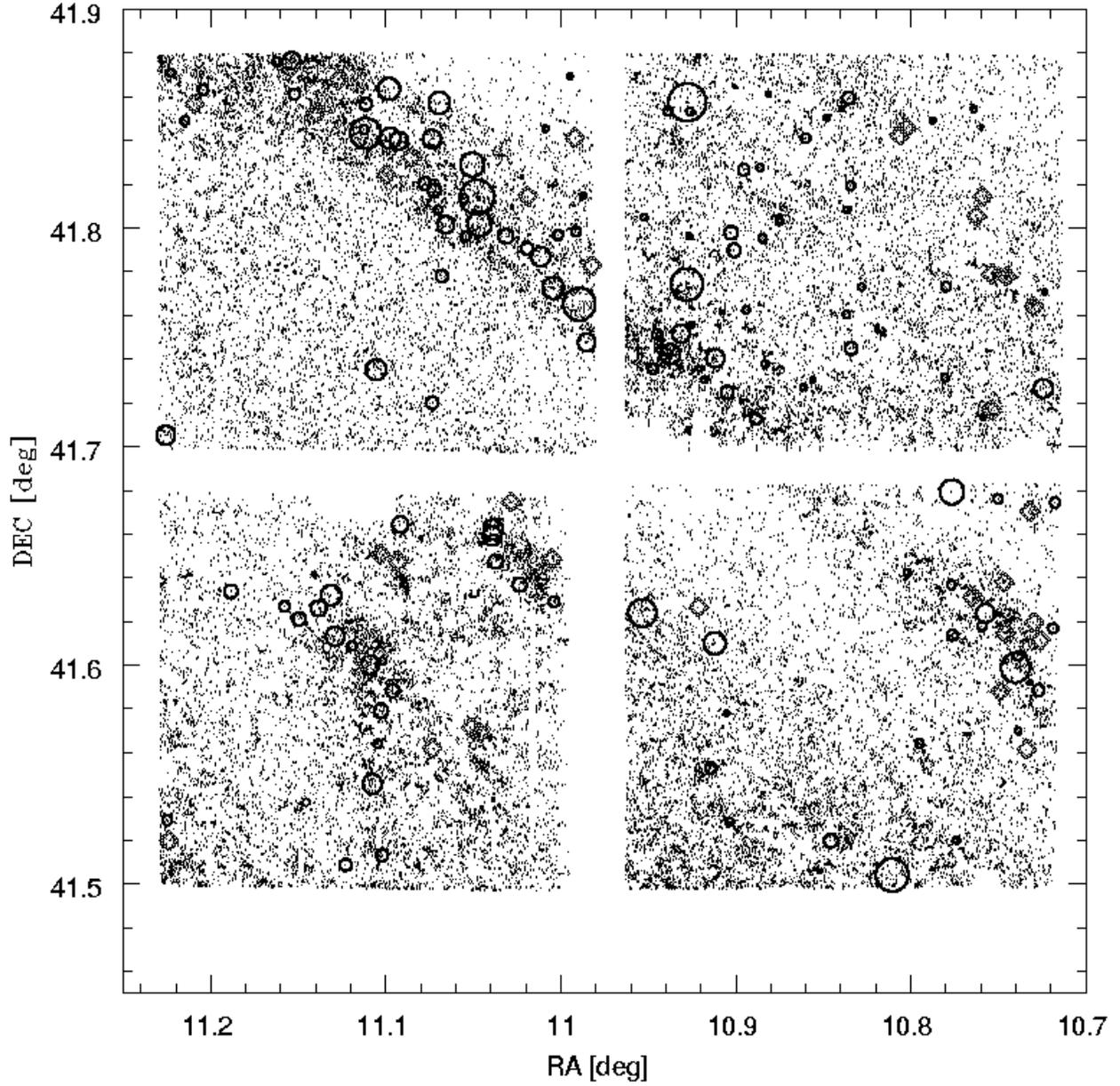}
\caption{Distribution of Cepheids and EBs in Field Y; symbols are
explained in previous figure. [{\it See the electronic edition of the
Journal for a color version of this figure.}]}
\label{ploty}
\end{figure}

Figure~\ref{vardist} shows the distribution of Cepheids found in all
the DIRECT M31 fields and Figure~\ref{ploty} shows Field Y
separately. The period of a Cepheid depends on its age, since its mass
determines the period of variability and the main sequence lifetime of
the star. Therefore, the distribution of Cepheids should trace star
formation and thus the spiral arms, which is the case. The longer
period or younger Cepheids appear to lie along the spiral arms. The
fact that the longest period of a Cepheid in M31Y is 25 days, while in
other fields there are 50 day period Cepheids, might indicate more
recent star formation on eastern side of the galaxy.

Table~\ref{totals} shows the number of each kind of variable found in
each DIRECT field. This paper thus almost doubles the DIRECT variables
in M31. Comparatively, Field M31Y is not as rich in variables as
Fields A, B and C which lie along the spiral arms, whereas it is
richer than Fields D and F.

To summarize, the observations of Field M31Y with the 4Shooter camera
on the FLWO 1.2~meter telescope resulted in the discovery of 41 EBs,
126 Cepheids, 48 other periodic variables and 49 miscellaneous
variables, almost doubling the number of M31 variables found by
DIRECT. Currently, there are 674 variables in M31 from the DIRECT
project: 89 EBs, 332 Cepheids and 253 other variables. Of the 264
variables in M31Y, 4 EBs and 20 Cepheids had been previously
discovered. We presented light curves and parameters for these
variables in this paper. A catalog of $BVI$ photometry of stars in
this field will be a subject of a future paper.

\acknowledgments{We thank the TAC of FLWO for the generous allocation
of the observing time. We would like to thank Perry Berlind, Saurabh
Jha, Jose Mu\~noz and Maria Contreras for obtaining observations for
this project. JK was supported by with NSF grant AST-9819787 and KBN
grant 5P03D004.21. Support for BJM and LMM was provided by NASA
through Hubble Fellowship grants HST-HF-01155.01-A and
HST-HF-01153.01-A, respectively, from the Space Telescope Science
Institute, which is operated by the Association of Universities for
Research in Astronomy, Incorporated, under NASA contract NAS5-26555.}

\begin{deluxetable}{lrrrrrrcrl}
\tabletypesize{\footnotesize}
\tablewidth{0pc}
\tablecaption{\sc DIRECT Eclipsing Binaries in M31Y}
\tablehead{
\colhead{} & \colhead{$P$} & \colhead{} & \colhead{} & \colhead{} &
\colhead{} & \colhead{} & \colhead{$i$} & \colhead{} & \colhead{} \\
\colhead{Name} & \colhead{$(days)$} & \colhead{$B_{max}$} & 
\colhead{$V_{max}$} & \colhead{$I_{max}$} & \colhead{$R_1$} & 
\colhead{$R_2$} & \colhead{(deg)} & \colhead{$e$} &
\colhead{Comments$^{\rm a}$}}
\startdata
D31J04422.4+413851.1 & 0.23278&  18.75 &17.84 &16.62 &  0.66 & 0.34 &  61.78 & 0.03 & V438, W UMa\\
D31J04420.5+414955.9 & 0.58390&  16.97 &16.67 &16.38 &  0.38 & 0.36 &  67.96 & 0.00 & \small Foreground\\
D31J04255.7+414013.0 & 0.83605&  21.03 &21.20 &20.23 &  0.65 & 0.34 &  66.48 & 0.00 & \\
D31J04301.0+414259.8 & 1.17912&  21.55 &21.44 &21.00 &  0.50 & 0.37 &  82.60 & 0.01 & \\
\enddata
\tablecomments{Table 4 is available in its entirety in the electronic
version of the Astronomical Journal. A portion is shown here for
guidance regarding its form and content.}
\label{tab:ebs}
\tablenotetext{a}{IDs of EBs found in M31B \citep{Kal98} and \citet{Mag97} are listed.}
\end{deluxetable}

\begin{small}
\begin{deluxetable}{ccccc}
\tabletypesize{\footnotesize}
\tablewidth{0pc}
\tablecaption{\sc Light Curves of Eclipsing Binaries in M31Y}
\tablehead{ \colhead{Name} & \colhead{Filter} &
\colhead{HJD$-$2450000} & \colhead{mag/flux} & \colhead{$\sigma_{mag}/\sigma_{flux}$} }
\startdata
D31J04256.2+413341.4......& B & 1379.9287  &   20.831 &  0.019\\ 
		    	  & B & 1428.8263  &   20.740 &  0.021\\ 
		       	  & B & 1429.8771  &   20.777 &  0.020\\ 
		          & B & 1432.8958  &   20.758 &  0.019\\ 
\enddata
\tablecomments{Table 3 is available in its entirety in the electronic
version of the Astronomical Journal. A portion is shown here for
guidance regarding its form and content.}
\label{tab:ecl_lc}
\end{deluxetable}
\end{small}

\begin{deluxetable}{lrrrrrl}
\tabletypesize{\footnotesize}
\tablewidth{0pc}
\tablecaption{\sc DIRECT Cepheids in M31Y}
\tablehead{
\colhead{} & \colhead{$P$} & \colhead{} & \colhead{} & \colhead{} &
\colhead{} & \colhead{} \\
\colhead{Name} & \colhead{$(days)$} & \colhead{$A_V$} &
\colhead{$\langle V\rangle$} & \colhead{$\langle I\rangle$} & \colhead{$\langle B\rangle$} &
\colhead{Comments}}
\startdata
D31J04255.7+413531.2 &  2.7600   & 0.28  & 22.00 & \nodata&  \nodata & \\
D31J04331.5+415140.9 &  2.8773   & 0.36  & 22.43 & \nodata&\nodata   & \\
D31J04302.1+414248.7 &  3.0787   & 0.46  & 22.00 & 21.41  &22.32     & \\
D31J04337.3+413441.1 &  3.1150   & 0.30  & 22.31 & 20.40  &23.05     & \\
\enddata                         
\tablecomments{Table 4 is available in its entirety in the electronic
version of the Astronomical Journal. A portion is shown here for
guidance regarding its form and content.}
\label{tab:ceph}
\end{deluxetable} 

\begin{small}
\begin{deluxetable}{ccccc}
\tabletypesize{\footnotesize}
\tablewidth{0pc}
\tablecaption{\sc Light Curves of Cepheids in M31Y}
\tablehead{ \colhead{Name} & \colhead{Filter} &
\colhead{HJD$-$2450000} & \colhead{mag/flux} & \colhead{$\sigma_{mag}/\sigma_{flux}$} }
\startdata
D31J04336.9+413141.8......& b& 1379.9287 &   -76.527&  97.018\\ 
 			  & b& 1428.8263 &  -456.584& 125.975\\ 
 			  & b& 1429.8771 &  -253.044& 111.878\\ 
		 	  & b& 1432.8958 &   381.949& 107.811\\ 

\enddata
\tablecomments{Table 5 is available in its entirety in the electronic
version of the Astronomical Journal. A portion is shown here for
guidance regarding its form and content.}
\label{tab:ceph_lc}
\end{deluxetable}
\end{small}

\begin{small}
\begin{deluxetable}{lrrrrr}
\tabletypesize{\footnotesize}
\tablewidth{0pc}
\tablecaption{\sc DIRECT Other Periodic Variables in M31Y}
\tablehead{
\colhead{} & \colhead{$P$} & \colhead{} & \colhead{} & \colhead{} \\
\colhead{Name} & \colhead{$(days)$} & \colhead{$\langle V \rangle$} & \colhead{$\langle I \rangle$} & \colhead{$\langle B \rangle$} & \colhead{Comments}}
\startdata
D31J04321.5+415124.2  &   1.8741 & 21.83 &   20.79  &  22.20 & \\
D31J04310.5+414331.0  &   3.0461 & 22.13 &   \nodata&  22.50 & \\
D31J04402.0+413212.7  &   3.6371 & 18.04 &   16.84  &  18.78 & \\
D31J04345.4+414850.5  &   4.1334 & 20.62 &   20.17  & \nodata& \\
\enddata
\tablecomments{Table 6 is available in its entirety in the electronic
version of the Astronomical Journal. A portion is shown here for
guidance regarding its form and content.}
\label{tab:per}
\end{deluxetable}
\end{small}

\begin{small}
\begin{deluxetable}{ccccc}
\tabletypesize{\footnotesize}
\tablewidth{0pc}
\tablecaption{\sc Light Curves of Other Periodic Variables in M31Y}
\tablehead{ \colhead{Name} & \colhead{Filter} &
\colhead{HJD$-$2450000} & \colhead{mag/flux} & \colhead{$\sigma_{mag}/\sigma_{flux}$} }
\startdata
D31J04308.0+413623.8......& b& 1379.9287 &   817.266&  88.453 \\
			  & b& 1428.8263 &   -66.370& 108.510 \\
			  & b& 1429.8771 &  -372.652&  99.494 \\
			  & b& 1432.8958 &   803.898&  94.021 \\
\enddata
\tablecomments{Table 7 is available in its entirety in the electronic
version of the Astronomical Journal. A portion is shown here for
guidance regarding its form and content.}
\label{tab:per_lc}
\end{deluxetable}
\end{small}

\begin{small}
\begin{deluxetable}{lrrrrrc}
\tabletypesize{\footnotesize}
\tablewidth{0pc}
\tablecaption{\sc DIRECT Miscellaneous Variables in M31Y}
\tablehead{
\colhead{Name} & \colhead{$\bar{V}$} & \colhead{$\bar{I}$} & 
\colhead{$\bar{B}$} & \colhead{Comments}}
\startdata
D31J04254.1+414033.9 &  19.75  &  16.37  & \nodata& \\ 
D31J04254.3+413219.4 &  23.27  &  19.43  & \nodata& \\
D31J04256.0+413509.2 &  18.62  &  17.52  &  18.87 & \\ 
D31J04257.6+413245.3 &  20.91  &  19.80  &  21.41 & \\ 
\enddata 	      		 
\tablecomments{Table 8 is available in its entirety in the electronic
version of the Astronomical Journal. A portion is shown here for
guidance regarding its form and content.}
\label{tab:misc}
\end{deluxetable}
\end{small}

\begin{small}
\begin{deluxetable}{ccccc}
\tabletypesize{\footnotesize}
\tablewidth{0pc}
\tablecaption{\sc Light Curves of Miscellaneous Variables in M31Y}
\tablehead{ \colhead{Name} & \colhead{Filter} &
\colhead{HJD$-$2450000} & \colhead{mag/flux} & \colhead{$\sigma_{mag}/\sigma_{flux}$} }
\startdata

D31J04254.3+413219.4......& b& 1379.9287 &    35.327 & 79.835  \\
			  & b& 1428.8263 &  -106.313 & 94.983  \\
			  & b& 1429.8771 &   -55.733 & 87.618  \\
			  & b& 1432.8958 &   -61.530 & 83.257  \\

\enddata
\tablecomments{Table 9 is available in its entirety in the electronic
version of the Astronomical Journal. A portion is shown here for
guidance regarding its form and content.}
\label{tab:misc_lc}
\end{deluxetable}
\end{small}

\begin{deluxetable}{lcccccr}
\footnotesize \tablewidth{11.cm} 
\tablecaption{Comparison of Variables in M31 DIRECT fields\label{totals}} \tablecolumns{6} 
\tablehead{ \colhead{Field} &  \colhead{Total} &  \colhead{} & \colhead{}&  \colhead{}&  \colhead{DIRECT} \\ \colhead{Name} &\colhead{Variables} &\colhead{EBs} &\colhead{Cepheids} &\colhead{Others} &\colhead{Paper} }
\startdata
M31A &  75 & 15 & 43 & 17 & Paper II \\
M31B &  85 & 12 & 38 & 35 & Paper I \\
M31C & 115 & 12 & 35 & 68 & Paper III \\
M31D &  71 &  5 & 38 & 28 & Paper IV \\
M31F &  64 &  4 & 52 &  8 & Paper V \\
M31Y & 264 & 41 &126 & 97 & Paper IX \\ 
\tableline
Totals &674& 89 &332 &253 & \\
\enddata
\end{deluxetable}


\clearpage
\begin{thebibliography}{99}

\bibitem[Alard \& Lupton(1998)]{Ala98} 
Alard, C., Lupton, R. 1998, \apj, 503, 325

\bibitem[Alard(2000)]{Ala00} 
Alard, C. 2000, A\&AS, 144, 363

\bibitem[Andersen(1991)]{And91} 
Andersen, J. 1991, A\&AR, 3, 91

\bibitem[Baade \& Swope(1963)]{Baa63}
Baade, W., Swope, H. H. 1963, \aj, 68, 435

\bibitem[Baade \& Swope(1965)]{Baa65}
Baade, W., Swope, H. H. 1965, \aj, 70, 212

\bibitem[Fitzpatrick et al.(2003)]{Fit03} 
Fitzpatrick, E.L., Ribas, I., Guinan E.F., Maloney, F.P., Claret, A
2003, \aj, in press (astro-ph/0301296)

\bibitem[Freedman \& Madore(1990)]{Fre90}
Freedman, W. L., Madore, B. F., 1990, \apj, 365, 186

\bibitem[Gaposchkin(1962)]{Gap62}
Gaposchkin, S. 1962, \aj, 67, 334

\bibitem[Guinan(1998)]{Gui98} 
Guinan, E.~F., et al.~1998, \apj, 509, L21

\bibitem[Harries, Hilditch \& Howarth(2003)]{Har03}
Harries, T. J., Hilditch, R. W., Howarth, I. D. 2003, \mnras, 339, 157

\bibitem[Hubble(1929)]{Hub29} 
Hubble, E. 1929, \apj, 69, 103

\bibitem[Joshi et al.(2003)]{Jos03} 
Joshi, Y.C., Pandey, A.K., Narasimha, D., Sagar, R., Giraud-Hiraud,
Y. 2003, A\&A, in press (astro-ph/0210373)

\bibitem[Kaluzny et al.(1998)]{Kal98} 
Kaluzny, J., Stanek, K.~Z., Krockenberger, M., Sasselov, D.~D., Tonry,
J.~L., Mateo, M. 1998, \aj, 115, 1016 (Paper I)

\bibitem[Kaluzny et al.(1999)]{Kal99} 
Kaluzny, J., Mochejska, B.~J., Stanek, K.~Z., Krockenberger, M.,
Sasselov, D.~D., Tonry, J.~L., Mateo, M. 1999, \aj, 118,346 (Paper IV)

\bibitem[Krockenberger et al.(1997)]{Kro97} 
Krockenberger, M., Sasselov, D.~D., Noyes, R. 1997, \apj, 479, 875

\bibitem[Landolt(1992)]{Lan92}
Landolt, A. 1992, \aj, 104, 340

\bibitem[Macri et al.(2001)]{Mac01} 
Macri, L.~M., Stanek, K.~Z., Sasselov, D.~D., Krockenberger, M.,
Kaluzny, J. 2001, \aj, 121, 870 (Paper VI)

\bibitem[Magnier et al.(1997)]{Mag97}
Magnier, E. A., Augusteijn, T., Prins, S., van Paradijs, J., Lewin,
W. H. G. 1997, A\&AS, 126, 401

\bibitem[Mochejska et al.(1999)]{Moc99} 
Mochejska, B.~J., Kaluzny, J., Stanek, K.~Z., Krockenberger, M.,
Sasselov, D.~D. 1999, \aj, 118, 2211 (Paper V)

\bibitem[Mochejska et al.(2001a)]{Moc01a} 
Mochejska, B.~J., Kaluzny, J., Stanek, K.~Z., Sasselov, D.~D.,
Szentgyorgyi, A.~H. 2001a, \aj, 121, 2032 (Paper VII)

\bibitem[Mochejska et al.(2001b)]{Moc01b} 
Mochejska, B.~J., Kaluzny, J., Stanek, K.~Z., Sasselov, D.~D.,
Szentgyorgyi, A.~H. 2001b, \aj, 121, 2032 (Paper VIII)

\bibitem[Monet(1996)]{Mon96}
Monet, D., et al. 1996, USNO-SA2.0 (Washington: US Naval Obs.)

\bibitem[Paczy\'nski(1997)]{Pac97}
Paczy\'nski, B. 1997, in The Extragalactic Distance Scale,
ed. M. Livio, M. Donahue \& N. Panagia (Cambridge: Cambridge
Univ.~Press), 273

\bibitem[Stanek et al.(1998a)]{Sta98a} 
Stanek, K.~Z., Kaluzny, J., Krockenberger, M., Sasselov, D.~D., Tonry,
J.~L., Mateo, M. 1998a, \aj, 115, 1894 (Paper II)

\bibitem[Stanek et al.(1998b)]{Sta98b} 
Stanek, K.~Z., Garnavich, P.M. 1998b, \apj, 503, 131

\bibitem[Stanek et al.(1999)]{Sta99} 
Stanek, K.~Z., Kaluzny, J., Krockenberger, M., Sasselov, D.~D., Tonry,
J.~L., Mateo, M. 1999, \aj, 117, 2810 (Paper III)

\bibitem[Szentgyorgyi et al.(2003)]{Sze03}
Szentgyorgyi, A.H. et al. 2003, in preparation

\bibitem[Walker(2003)]{Wal03}
Walker, A.R., to appear in ``Stellar Candles for the Extragalactic
Distance Scale'', Lecture Notes in Physics (Springer), ed. D. Alloin
and W. Gieren (astro-ph/0303011)

\bibitem[Welch et al.(1986)]{Wel86} 
Welch, D., L., McAlary, C. W., McLaren, R. A., Madore, B., F. 1986,
\apj, 305, 583


\end{thebibliography}
\end{document}